\documentclass[aps,prx,twocolumn,superscriptaddress,showpacs,floatfix]{revtex4-1}
\usepackage{amsmath,amssymb,amsfonts,amsthm}
\usepackage{graphicx}
\usepackage{bm}
\usepackage{bbm}
\usepackage{color}
\usepackage{dcolumn}   % needed for some tables
\usepackage{epstopdf}
\usepackage[caption=false]{subfig}
\usepackage{xr}
\usepackage{color}
\usepackage[colorlinks=true,linkcolor=blue,urlcolor=blue,citecolor=blue]{hyperref}

\begin{document}

 \newcommand{\breite}{1.0} %  for twocolumn

\newtheorem{prop}{Proposition}
\newtheorem{cor}{Corollary} 

\newcommand{\be}{\begin{equation}}
\newcommand{\ee}{\end{equation}}

\newcommand{\bea}{\begin{eqnarray}}
\newcommand{\eea}{\end{eqnarray}}
\newcommand{\lt}{<}
\newcommand{\gt}{>} 

\newcommand{\Reals}{\mathbb{R}}     % Reals
\newcommand{\Com}{\mathbb{C}}       % Complex #
\newcommand{\Nat}{\mathbb{N}}       % Natural #

\newcommand{\id}{\mathbboldsymbol{1}}    

\newcommand{\Real}{\mathop{\mathrm{Re}}}
\newcommand{\Imag}{\mathop{\mathrm{Im}}}

\def\O{\mbox{$\mathcal{O}$}}   % Order epsilon ... 
\def\F{\mathcal{F}}			% FourierTrafo
\def\sgn{\text{sgn}}

\newcommand{\deo}{\ensuremath{\Delta_0}}
\newcommand{\dea}{\ensuremath{\Delta}}
\newcommand{\ak}{\ensuremath{a_k}}
\newcommand{\ad}{\ensuremath{a^{\dagger}_{-k}}}
\newcommand{\sx}{\ensuremath{\sigma_x}}
\newcommand{\sz}{\ensuremath{\sigma_z}}
\newcommand{\spl}{\ensuremath{\sigma_{+}}}
\newcommand{\smi}{\ensuremath{\sigma_{-}}}
\newcommand{\alk}{\ensuremath{\alpha_{k}}}
\newcommand{\bk}{\ensuremath{\beta_{k}}}
\newcommand{\ok}{\ensuremath{\omega_{k}}}
\newcommand{\vd}{\ensuremath{V^{\dagger}_1}}
\newcommand{\vi}{\ensuremath{V_1}}
\newcommand{\vo}{\ensuremath{V_o}}
\newcommand{\zc}{\ensuremath{\frac{E_z}{E}}}
\newcommand{\xc}{\ensuremath{\frac{\Delta}{E}}}
\newcommand{\xd}{\ensuremath{X^{\dagger}}}
\newcommand{\aok}{\ensuremath{\frac{\alk}{\ok}}}
\newcommand{\tpw}{\ensuremath{e^{i \ok s }}}
\newcommand{\tpe}{\ensuremath{e^{2iE s }}}
\newcommand{\tmw}{\ensuremath{e^{-i \ok s }}}
\newcommand{\tme}{\ensuremath{e^{-2iE s }}}
\newcommand{\epls}{\ensuremath{e^{F(s)}}}
\newcommand{\emis}{\ensuremath{e^{-F(s)}}}
\newcommand{\epl}{\ensuremath{e^{F(0)}}}
\newcommand{\emi}{\ensuremath{e^{F(0)}}}

\newcommand{\lr}[1]{\left( #1 \right)}
\newcommand{\lrs}[1]{\left( #1 \right)^2}
\newcommand{\lrb}[1]{\left< #1\right>}
\newcommand{\nbt}{\ensuremath{\lr{ \lr{n_k + 1} \tmw + n_k \tpw  }}}

\newcommand{\om}{\ensuremath{\omega}}
\newcommand{\dw}{\ensuremath{\Delta_0}}
\newcommand{\wbp}{\ensuremath{\omega_0}}
\newcommand{\dv}{\ensuremath{\Delta_0}}
\newcommand{\vbp}{\ensuremath{\nu_0}}
\newcommand{\vplus}{\ensuremath{\nu_{+}}}
\newcommand{\vminus}{\ensuremath{\nu_{-}}}
\newcommand{\wplus}{\ensuremath{\omega_{+}}}
\newcommand{\wminus}{\ensuremath{\omega_{-}}}
\newcommand{\uv}[1]{\ensuremath{\mathbf{\hat{#1}}}} % for unit vector
\newcommand{\abs}[1]{\left| #1 \right|} % for absolute value
\newcommand{\avg}[1]{\left< #1 \right>} % for average
\let\underdot=\d % rename builtin command \d{} to \underdot{}
\renewcommand{\d}[2]{\frac{d #1}{d #2}} % for derivatives
\newcommand{\dd}[2]{\frac{d^2 #1}{d #2^2}} % for double derivatives
\newcommand{\pd}[2]{\frac{\partial #1}{\partial #2}} 
% for partial derivatives
\newcommand{\pdd}[2]{\frac{\partial^2 #1}{\partial #2^2}} 
% for double partial derivatives
\newcommand{\pdc}[3]{\left( \frac{\partial #1}{\partial #2}
 \right)_{#3}} % for thermodynamic partial derivatives
\newcommand{\ket}[1]{\left| #1 \right>} % for Dirac bras
\newcommand{\bra}[1]{\left< #1 \right|} % for Dirac kets
\newcommand{\braket}[2]{\left< #1 \vphantom{#2} \right|
 \left. #2 \vphantom{#1} \right>} % for Dirac brackets
\newcommand{\matrixel}[3]{\left< #1 \vphantom{#2#3} \right|
 #2 \left| #3 \vphantom{#1#2} \right>} % for Dirac matrix elements
\newcommand{\grad}[1]{{\nabla} {#1}} % for gradient
\let\divsymb=\div % rename builtin command \div to \divsymb
\renewcommand{\div}[1]{{\nabla} \cdot \boldsymbol{#1}} % for divergence
\newcommand{\curl}[1]{{\nabla} \times \boldsymbol{#1}} % for curl
\newcommand{\laplace}[1]{\nabla^2 \boldsymbol{#1}}
\newcommand{\vs}[1]{\boldsymbol{#1}}
\let\baraccent=\= % rename builtin command \= to \baraccent
%%%%%%%%%%%%%%%%%%%%%%%%%%%%%%%%%%%%%%%%%%%%%
% End Definitions
%%%%%%%%%%%%%%%%%%%%%%%%%%%%%%%%%%%%%%%%%%%%%

%Title of paper
\title{{\em Draiding} Majoranas to dynamically engineer Hamiltonians}

\author{Ivar Martin}
\email{ivar@anl.gov}
\affiliation{Material Science Division, Argonne National Laboratory, Argonne, IL 08540, USA}
\author{Kartiek Agarwal}
\email{agarwal@physics.mcgill.ca}
\affiliation{Department of Physics, McGill University, Montr\'{e}al, Qu\'{e}bec H3A 2T8, Canada}

\date{\today}
\begin{abstract}

We propose and  analyze a family of periodic braiding protocols in systems with multiple localized Majorana modes ({\em majoranas}) for the purposes of Hamiltonian engineering. The protocols rely on  double braids -- {\em draids} -- which flip the signs of both majoranas, as one is taken all the way around the other. Rapid draiding dynamically suppresses some or all inter-majorana couplings. Protocols suppressing all couplings can drastically reduce residual dynamics within the nearly degenerate many-body subspace, producing more robust computational subspace. Non-trivial topological models can be achieved by selectively applying draids to some of overlapping (imperfect) majoranas. Importantly, draids can be implemented without having to physically braid majoranas or using projective measurements. In particular, draids can be performed by periodically modulating the coupling between a quantum dot and topological superconducting wire to dynamically suppress the hybridization of majoranas by more than an order of magnitude in current experimental setups. 

\end{abstract}
\maketitle

\section{Introduction}

Isolated Majorana fermions (or {\em majoranas} for short) play a key role in a number of proposals for topological quantum computing. Any conventional fermion can always be expanded in terms of two self-adjoint majorana operators, $c^\dagger  = \gamma_1 + i\gamma_2$, which satisfy the algebra $\{\gamma_i, \gamma_j\} = 2\delta_{ij}$. However, in topological superconductors,  majoranas can exist in spatial separation from each other~\cite{kitaev2001unpaired,readgreenmajorana,fu2008superconducting,saumajorana,lutchysausarma,oregrefaelvonoppen}. Every pair of decoupled majoranas increases the ground state degeneracy by a factor of two, implementing a Hilbert subspace that is protected from decoherence due to local noise.  Majoranas can further be braided to realize nontrivial operations within this multidimensional subspace~\cite{ivanov_majorana,alicea2011non,pankratova2018multi}.

Unfortunately, true degeneracy in this subspace
is only achieved when majoranas are perfectly isolated from one another spatially. In any realistic implementation~\cite{mourik2012signatures,rokhinson2012fractional,deng2012anomalous,das2012zero,finck2013,churchill2013,nadj2014observation,albrecht2016exponential,zhang2018quantized,wang2018evidence} (see Ref.~\cite{lutchyn2018majorana} for a more complete set of references), a finite spatial overlap between majoranas lifts degeneracy, and diminshes the time in which the encoded quantum state decoheres. Thus, finding ways to reduce the majorana overlaps is desirable in order to bring topological quantum computing (TQC) closer to reality.
 
In this work, we demonstrate how majoranas can be dynamically decoupled by means of periodic braiding, thereby improving the quality (longevity) of quantum states in the ground state manifold. Central to our protocol is the role of double braids, henceforth termed \emph{draids}, which do not change majorana locations, but instead administer a $\pi$-phase to them. Using this sign flip, which is of a robust, topological origin, dynamical decoupling can be performed provided the braiding frequency is faster than the largest overlap between any two majoranas in the  system. 

In addition to nullifying a generic Hamiltonian of interacting majoranas as a whole, it is also possible to selectively {\em trim} the Hamiltonian. Thus, the approach may be also used towards dynamically realizing models of interacting majorana fermions that have generated theoretical interest recently from Sachdev-Ye-Kitaev type models~\cite{maldacenacomments,pikulinfranz,chewsyk,jaewonlowranksyk,etiennefamilyofsyk}, majorana-Hubbard models~\cite{affleckmajoranahubbard,rahmanimajoranahubbardladder}, and topological models~\cite{kitaev2006anyons,yaokivelson,sagimajoranaspinliquid}. Our approach here is a departure from previous approaches in which majoranas are arranged carefully in special geometries in order to realize certain unconventional Hamiltonians~\cite{barkeshli2015physical,landau2016,plugge2016,karzighexon2017,romito2017,sagimajoranaspinliquid,oreg2020majorana}. In those approaches, the locality of majoranas is used to arrive at Hamiltonians where only certain terms are significant in magnitude. Our approach places less stringent demands on the specific geometry in which majoranas are arranged, focusing instead on relative enhancing of desirable  terms by reducing the amplitude of offending ones using braiding protocols. 

The dynamical decoupling protocols we describe combine multiple draids occurring at different multiples of a single frequency, in the framework of `polyfractal' decoupling schemes, originally proposed the context of few spin systems~\cite{khodjastehlidarviolaPRL1,lidarhamiltoniancancel}, and extended by the authors to many-body systems to dynamically create novel symmetries~\cite{agarwal2019polyfractal}. 
While in general the protocol results in deleterious heating of many-body systems~\cite{kuwahara2016floquet,abanin2017rigorous,MoriSaitoKuwahara,elseprethermal}, it only occurs on an exponentially long time scale provided that the braiding is performed at a frequency faster than majorana overlaps. Under these conditions, an effective local Hamiltonian description applies for extended times and an appropriate choice of a periodic braiding scheme can be used to engineer a wide variety of  effective Hamiltonians. We describe some such schemes applied to  a generic starting Hamiltonian operating on a set of localized majorana modes (with finite non-zero overlaps) to construct the simple Kitaev chain with localized edge majoranas and the $Z_2 \times Z_2$ symmetry-protected topological insulator. We also provide numerical support for these claims.

To make connection with  experiments, we present a scheme to improve the quality of the zero-bias peak in the current generation of experiments on superconducting wires that host nearly isolated majorana fermions. Our protocol involves performing repeated Landau-Zener tunneling between a quantum dot housing a single fermionic mode, and the superconducting wire, by varying the local potential of the quantum dot in time. As we show, this protocol may be interpreted as performing repeated draids between one majorana each from the quantum dot and the wire. We show that the scheme can lead to more than an order of magnitude suppression of the hybridization between the majoranas in quantum wires in current experiments. This example serves as an illustration how draids can be implemented without having to physically braid majoranas or performing projective measurements~\cite{bonderson2008measurement,vijay2016teleportation}. 

\begin{figure}[htp]
\includegraphics[width=2.5in]{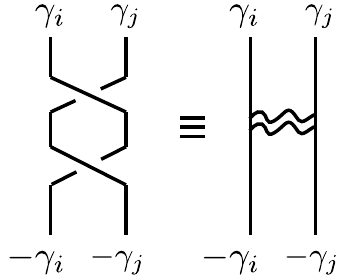}
\caption{The double braid or draid can be used to change the sign of a pair of majoranas.}
\label{fig:elementary}
\end{figure}
This paper is organized as follows. In Sec.~\ref{sec:elementary}, we discuss certain elementary protocols using a few majoranas that will form the basis of more complex decoupling schemes on many-body majorana Hamiltonians which we will describe later. In Sec.~\ref{sec:polyfrac}, we summarize the polyfractal braiding protocol, which was introduced in our previous work Ref.~\cite{agarwal2019polyfractal}, and the results necessary for the following discussion. In Sec.~\ref{sec:Hcancel}, we describe and numerically test protocols for improving the degeneracy of the Hilbert space starting from a set of overlapping, interacting majorana fermions. In Sec.~\ref{sec:Honed}, we work with one-dimensional arrays of majoranas, and discuss protocols for creating novel topological phases in these systems. We provide numerical support using exact-diagonalization studies. Finally, in Sec.~\ref{sec:Hexp}, we discuss a specific implementation of our ideas to improve the majorana zero-bias peak seen in the current generation of experiments.  

\section{Elementary operation}
\label{sec:elementary}

The elementary schemes that we consider below are motivated by the concept of dynamical decoupling~\cite{slichter2013principles}, which  has its origins in nuclear magnetic resonance (NMR). The aim there is to remove unwanted (say, inhomogeneous broadening) features in spectral data using appropriate magnetic pulses (for instance, using spin echo protocols). Translated into the language of Hamiltonian engineering, in its simplest form, these pulses implement dynamical evolution via the physical Hamiltonian $H = A + B$ for half a period of time, and a related Hamiltonian $H' = A - B$ for the other half period. When this period is sufficiently short, the effective Hamiltonian describing the dynamics on much longer time scales can be expected to be described by the average $H_{\text{eff}} \approx (H + H')/2 = A$ modulo $\mathcal{O} \left( T \right)$ corrections, where $T$ is the time period. In this way, one can get rid of undesirable (in this case, $B$) terms in the Hamiltonian. (Precisely how this works for many-body systems in discussed in Sec.~\ref{sec:polyfrac}.)

To perform such a sign-reversal operation with Hamiltonians of majorana operators, we require flipping signs of either individual, or pairs of majoranas \emph{without changing their spatial configuration}. The latter turns out to be rather easy to achieve with majoranas. A pair of majoranas can be braided with one another to robustly implement the relation $\left( \gamma_i , \gamma_j \right) \rightarrow \left( \gamma_j, -\gamma_i \right)$. By performing this sequence twice, one naturally finds

\be
\left( \gamma_i , \gamma_j \right) \xrightarrow{\text{Double Braid / Draid}} \left( - \gamma_i, - \gamma_j \right).
\ee

This operation is graphically illustrated in Fig.~\ref{fig:elementary}.  The advantage of performing such an operation is that it is topologically robust---the $\pi-$ phase shift has a topological origin which is thus robust to imperfections in the braiding protocol. Since these double braids are central to the protocols we discuss below, we term these \emph{draids} for notational convenience.   

(Note that there still is the subtle issue of making errors such as being unable to bring the majoranas back to their original position at the end of the protocol, but such errors are subject to averaging out over the course of many cycles, presuming there is no systematic offset in the implementation of braiding.) 

Any dynamical phase accrued in this process may be considered small as long as the braiding is performed on a timescale much faster than the overlap between the majoranas. Note that this property is special to majorana fermions (or other realizations of Ising non-Abelian anyons); two complex fermions will not accrue any (statistical/topological) phase in the process of being braided twice and returned to their original locations. 

\section{Summary of the polyfractal protocol}
\label{sec:polyfrac}

\begin{figure}[htp]
\includegraphics[width=2.5in]{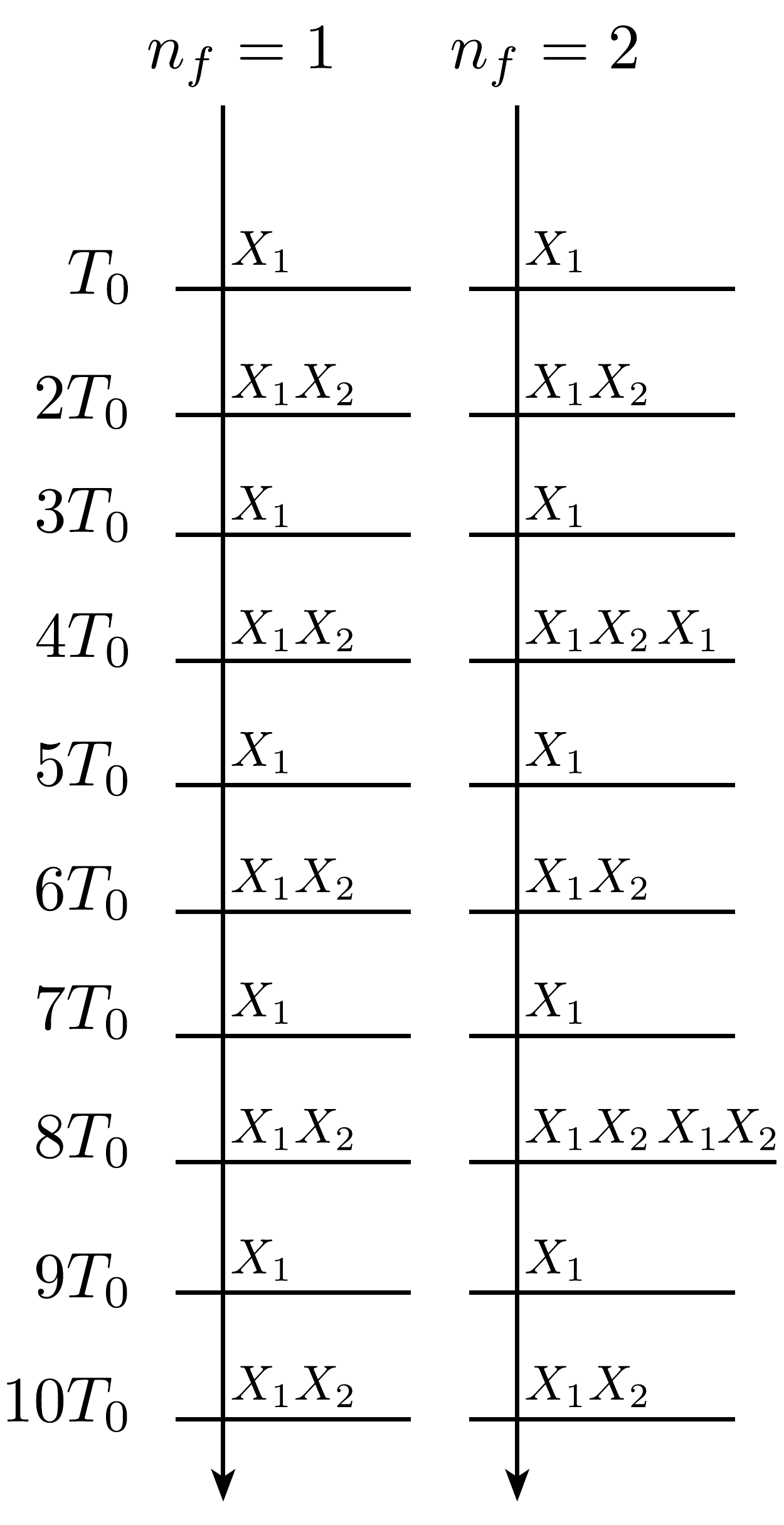}
\caption{Illustration of the polyfractal protocol used for dynamical decoupling. The protocol involves unitary evolution under the system's physical Hamiltonian, punctuated by the application of operators $X_i$ at specific times. The above illustrates when these operators are applied for the case where the number of operators, $n_s = 2$, and number of fractal layers, $n_f = 1,2$ up to $t = 10 T_0$ (note that the period for the protocol at $n_f = 2$ is $16 T_0$; thus the entire period is not illustrated).  Note that at times where the same $X_i$ appears an even number of times, these operators can be ignored. Further, in cases where $X_i X_j$ is applied, but $X_i$ and $X_j$ share majoranas, the actual braids required to be performed can be simplified---for instance, the majoranas that appear in both $X_i$ and $X_j$ need not be braided.}
\label{fig:polyfrac}
\end{figure}

Multiple distinct elementary braiding operations can be combined and used to selectively eliminate terms in majorana Hamiltonians. Before we describe some such protocols, we comment on the general theory underpinning dynamical decoupling using `polyfractal protocols', introduced by the present authors in Ref.~\cite{agarwal2019polyfractal}. 

In what follows, we denote a set of multiple simultaneous draiding operations, between non-overlapping pairs of majoranas, by the unitary $X_i$, that satisfies $X^2_i = 1$. For instance, $X_i$ may correspond to draiding majoranas $1$ and $2$, and $5$ and $6$, and be therefore given by the product $X_i = \gamma_1 \gamma_2 \gamma_5 \gamma_6$. 

The polyfractal protocol 
corresponds to time evolution of the system under its physical Hamiltonian $H$, but punctuated by the application of the operations $X_i$ periodically with increasingly-longer periods related by factors of $2$. Note that we aim for the unique set of operators $X_i$, $i \in [1,n_s]$ to be symmetry generators of the effective Hamiltonian obtained from this dynamical scheme. Periodic application of a particular $X_i$ allows one to suppress terms in the effective Hamiltonian that do not commute with $X_i$. In the same spirit, by applying multiple $X_i$ at periods $T_i = 2^{i T_0}$, we can make sure that the resultant Hamiltonian commutes with all $X_i$. Next, the dynamical suppression of non-commuting terms can be further enhanced by repeating this entire sequence at twice the period, and this represents a second `fractal layer' of protocols. In total, we imagine applying the unitary $X_i$ at period $T = 2^{i + j n_s} T_0$, where $i \in \{0,1,...,n_s -1\}$, $j \in \{0,1,...,n_f -1 \}$, and $n_s$ and $n_f$ denote the total number of distinct operators $X_i$, and the number of fractal layers, respectively. 

A particular illustration of the times at which these operators are applied is exhibited in Fig.~\ref{fig:polyfrac} for $n_s = 2$, but different fractal layers $n_f = 1,2$. Concretely, for $n_f = 1$, the operator $X_1$ is applied with period $T_0$, and $X_2$ with period $2T_0$; Ref.~\cite{agarwal2019polyfractal} shows that an effective Hamiltonian then describes the dynamics stroboscopically at periods $T_F = 4 T_0$~\footnote{Note that we look at the Floquet unitary at twice the period of the slowest drive in the protocol. This is related to the fact that only after an application of the operator $X_i$ twice, does an effective Hamiltonian that approximately commutes with $X_i$ emerge.}, and importantly, commutes with $X_1, X_2$ up to terms of order $\mathcal{O} \left( T_0 h \right)$, where $h$ is the typical overlap between majoranas. For $n_f = 2$, the operators $X_1, X_2$ are reapplied, now at periods $4T_0, 8T_0$, respectively. The effective Hamiltonian in this case, that describes dynamics stroboscopically at the Floquet period $T_F = 16 T_0$, commutes with both $X_1, X_2$ to order $\mathcal{O} \left( (T_0 h)^2 \right)$. 

This process cannot be continued indefinitely because that involves driving the system at smaller and smaller frequencies, and eventually leads to  heating. In particular, there exists an optimal number of fractal layers given by $n^*_f \sim \text{log} \left( \frac{1}{n_s} \text{log}_\text{2} \left( \frac{1}{T_0 h} \right) \right)$.  For this  choice of $n_f$, the heating rate  is stretched-exponentially long in $1/(T_0 h)$; thus heating may be ignored for a long time within this scheme, while the time up to which the effective Hamiltonian may be assumed to commute with $X_i$, scales as $t_* \sim h \left( \frac{1}{T_0 h} \right)^{n^*_f}$, which grows polynomially in the drive frequency, but with a power that itself grows as $T_0 \rightarrow 0$.

The commutation of operators $X_i$ with the effective Hamiltonian precludes the presence of certain terms. Thus, the polyfractal protocol provides one way of eliminating a large set of such terms dynamically, in a way analogous to dynamical decoupling methods used in NMR. With an appropriate choice of effective symmetry generators $X_i$, we may dramatically trim the effective majorana Hamiltonian as per our requirements, and we explore some applications of this next.

\section{Hamiltonian Cancellation}
\label{sec:Hcancel}
\begin{figure}[htp]
\includegraphics[width=3.1in]{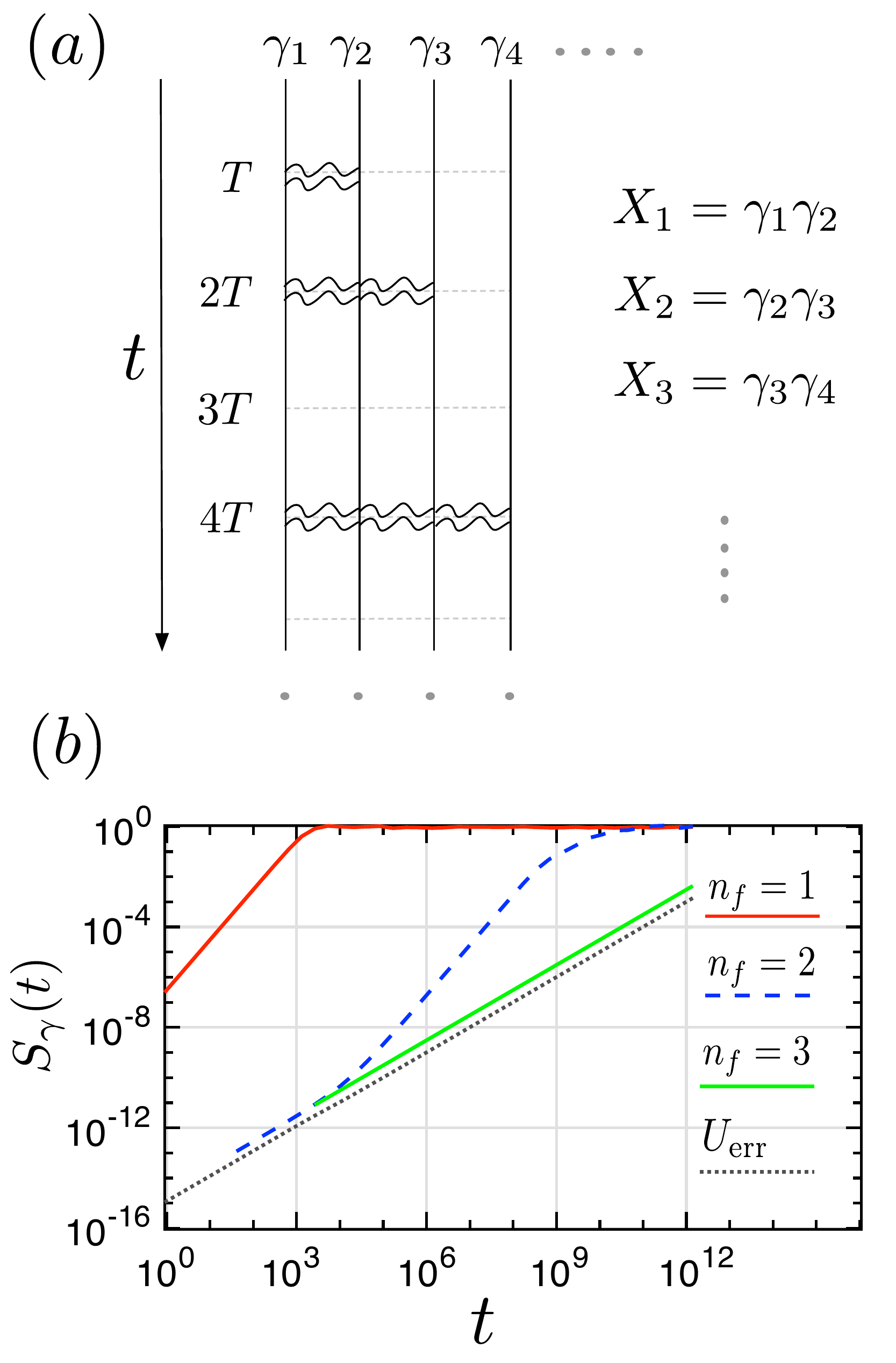}
\caption{(a) Protocol for cancellation of a generic Hamiltonian of majoranas. Note that the draids at time $4 T$, for instance, may be simplified to a single draid between majoranas $1, 4$. (b) Relaxation of coherence of all majorana pairs in time,  probed for a system of 8 majoranas, and averaged over 100 disorder realizations. We show only the majorana pair which relaxes the fastest in time for a given realization of the protocol, denoted by the number of fractal layers, $n_f = 1,2,3$. In particular, we plot $S_\gamma (t) = \text{max}_{ij} \left[ 1 - \langle \gamma_i (t) \gamma_j (t) \gamma_j \gamma_i \rangle \right]$, which approaches $1$ when the (fastest relaxing) majorana pair operator $\gamma_i \gamma_j$ relaxes. $U_{\text{err}}$ is also plotted and denotes is the Frobenius-norm error in the unitarity of the time-evolution matrix; this error provides an effective lower bound to the relaxation amplitudes defined above as it occurs simply due to numerical inaccuracy in computing the time-evolution matrix at exponentially long times. The results exhibit strong suppression of the Hamiltonian as the number of fractal layers is increased. Here the fastest drive period is $T_0 = 0.01$, the slowest drive period is $2^{6 n_f - 1}T_0$, and the effective Hamiltonian describes dynamics stroboscopically at period of $T_F = 2^{6 n_f} T_0$.}
\label{fig:Hcancel}
\end{figure}

As a first application of the polyfractal protocols, we consider the full nullification of the Majorana Hamiltonian, within a given parity sector. 
To cancel the full Hamiltonian, modulo the parity $P = \Pi_i \sqrt{i} \gamma_i$, we need to generate symmetry under draids between all possible majorana pairs. To see this, first suppose there is a term in the Hamiltonian that contains $\gamma_n$ but not $\gamma_m$. A draid exchanging $\gamma_n$ and $\gamma_m$ changes the sign of such a term. Then, as per Ref.~\cite{agarwal2019polyfractal}, applying the draid operator operator $X_i = i \gamma_n \gamma_m$ in a fractal fashion will eliminate such a term. Continuing this logic, if we generate symmetry with respect to all possible majorana pairs, the Hamiltonian cannot have any term except a term proportional to the parity. 

To generate symmetry with respect of all majorana pairs, it suffices to draid $N-2$ of possible $N-1$ ``nearest-neighbor" majoranas described by the unitary operator $X_i = i \gamma_i \gamma_{i+1}$. Note that we need $N-2$ generators, because the majorana pair left out is equivalent to a product of all other pairs multiplied by the parity $P$. Implementing all these generators leads to the cancellation of all terms in the Hamiltonian, which subsequently tends to zero as $T_0\to 0$. 

The efficacy of the protocol in canceling all terms can be gauged by examining the timescale at which majorana pairs, which can be viewed as conventional complex fermions, relax. In particular, we examine the parity $S_{ij} (t) = 1 - \avg{\gamma_i (t) \gamma_j (t) \gamma_j \gamma_i}$, for all possible majorana pairs $(\gamma_i, \gamma_j)$, and where the averaging is performed over all initial states. Note that $S_{ij}(t) \rightarrow 0$ at $t \rightarrow 0$, indicating the parity operator retains information about the initial state at $t = 0$ (trivially). The timescale at which the operator relaxes is approximately given by the time when $S_{ij}(t)$ approaches its equilibrium value $\sim 1/(N/2)$---note that the global parity is a conserved quantity, and the local parity thus behaves as a conserved density which relaxes by spreading out throughout the system. In Fig.~(\ref{fig:Hcancel}), we examine the efficacy of the protocol by plotting the fastest relaxing $S_{ij}(t)$ for a given implementation of the polyfractal protocol for a given fractal layer number, $n_f$. The protocol clearly becomes more effective as $n_f$ is increased between 1 and 3. (In these simulations, we consider the most general parity conserving majorana  Hamiltonian constructed out of 8 majoranas; we do so by generating random matrices in the parity-fixed Hilbert space with each element being a complex number whose real and imaginary parts are drawn from a uniform distribution in $[-1,1]$.)

\section{One-dimensional models}
\label{sec:Honed}
\begin{figure}[htp]
\includegraphics[width=3.2in]{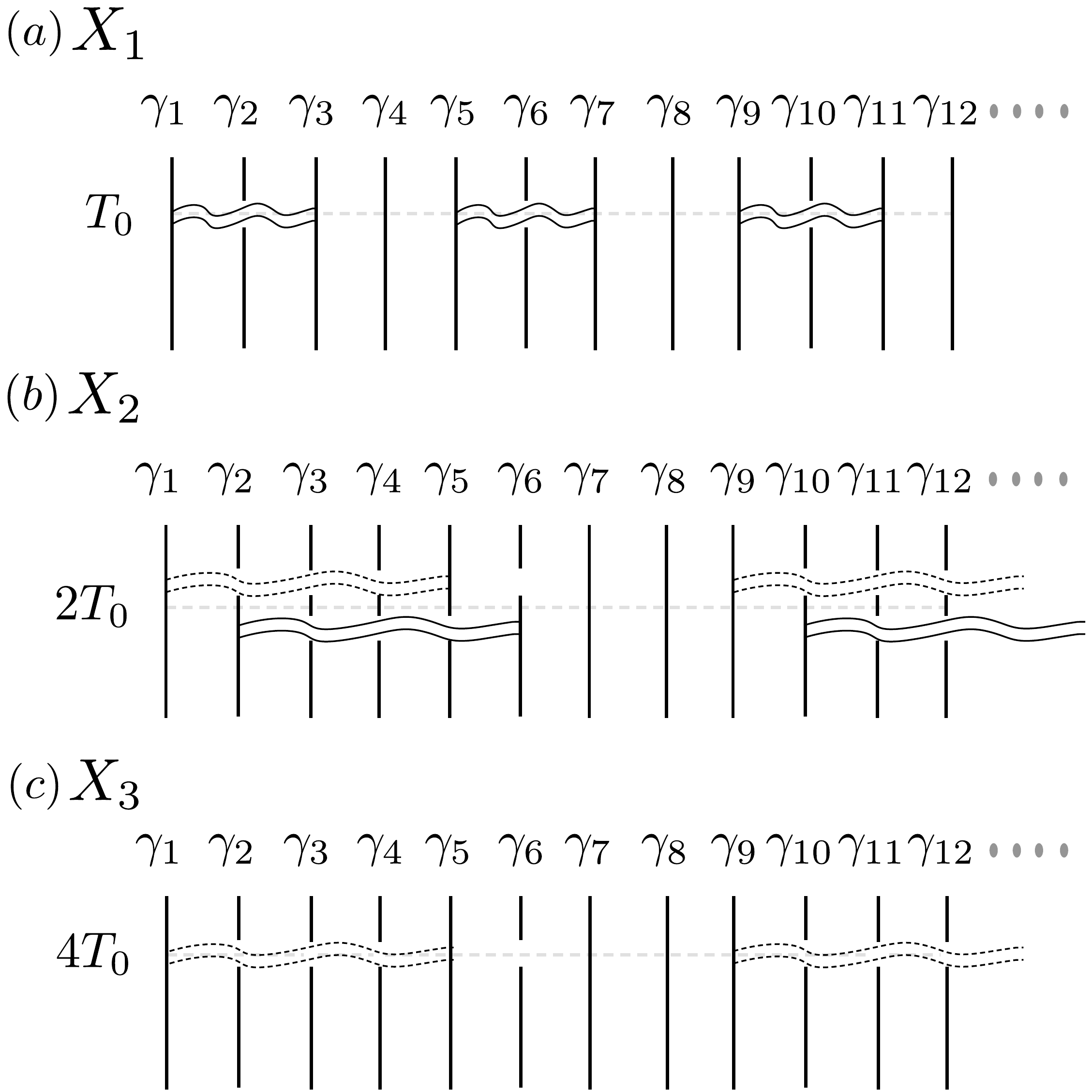}
\caption{(a) Operator $X_1$ applied every $T_0$ involves braiding every other majorana in a majorana chain. This leads to the cancelation of nearest-neighbor tunneling. The operation may be thought of as effectively splitting the chain into two chains, of exclusively the odd-numbered (even-numbered) majoranas. (b) Operator $X_2$ applied every $2 T_0$ involves braiding every other majorana in each of the two split chains. This operations leads to canceling tunneling upto distance $3$. (c) Operator $X_3$ applied every $4 T_0$ can be used to cancel nearest neighbor majorana interactions. Finally, note that to simplify the protocol, one may choose $X_2$ to be instead given by $X_2 X_3$ which reduces the number of braiding operations every $2T_0$ while achieving the same result in terms of cancelation.}
\label{fig:elimoned}
\end{figure}

\subsection{Hamiltonian suppression}
The scheme in the previous section cancels any Majorana Hamiltonian, but takes an exponentially long time in the system size $N$ (the time period $T_F = 2^{n_f(N-2)} T_0$). On the other hand, if the Hamiltonian has some special structure, cancellation can be done more efficiently. For instance, if the majoranas are arranged in a line, only a small subset of terms in Hamiltonian have large magnitude, and these can be canceled more efficiently.  

For instance, if only nearest-neighbor hopping is significant, then it can be suppressed in just one step everywhere in the chain  by braiding pairs of majoranas in odd positions with one another at time period $T_0$ (Fig.~\ref{fig:elimoned} a). One can perform the same operation within the two split chains at periods of $2 T_0 $, to suppress majorana tunneling up to distance $2^2-1$ (Fig.~\ref{fig:elimoned} b). In general, $2^n-1$-range hopping terms may be eliminated at time $2^n T_0$. Interactions involving four majoranas can be eliminated similarly---if for instance, interactions involving four adjacent majoranas are most relevant, such terms may be eliminated by braiding alternate majoranas in only one of the split chains (Fig.~\ref{fig:elimoned} c). 

\subsection{1d topological phases}
\begin{figure}[htp]
\includegraphics[width=3.1in]{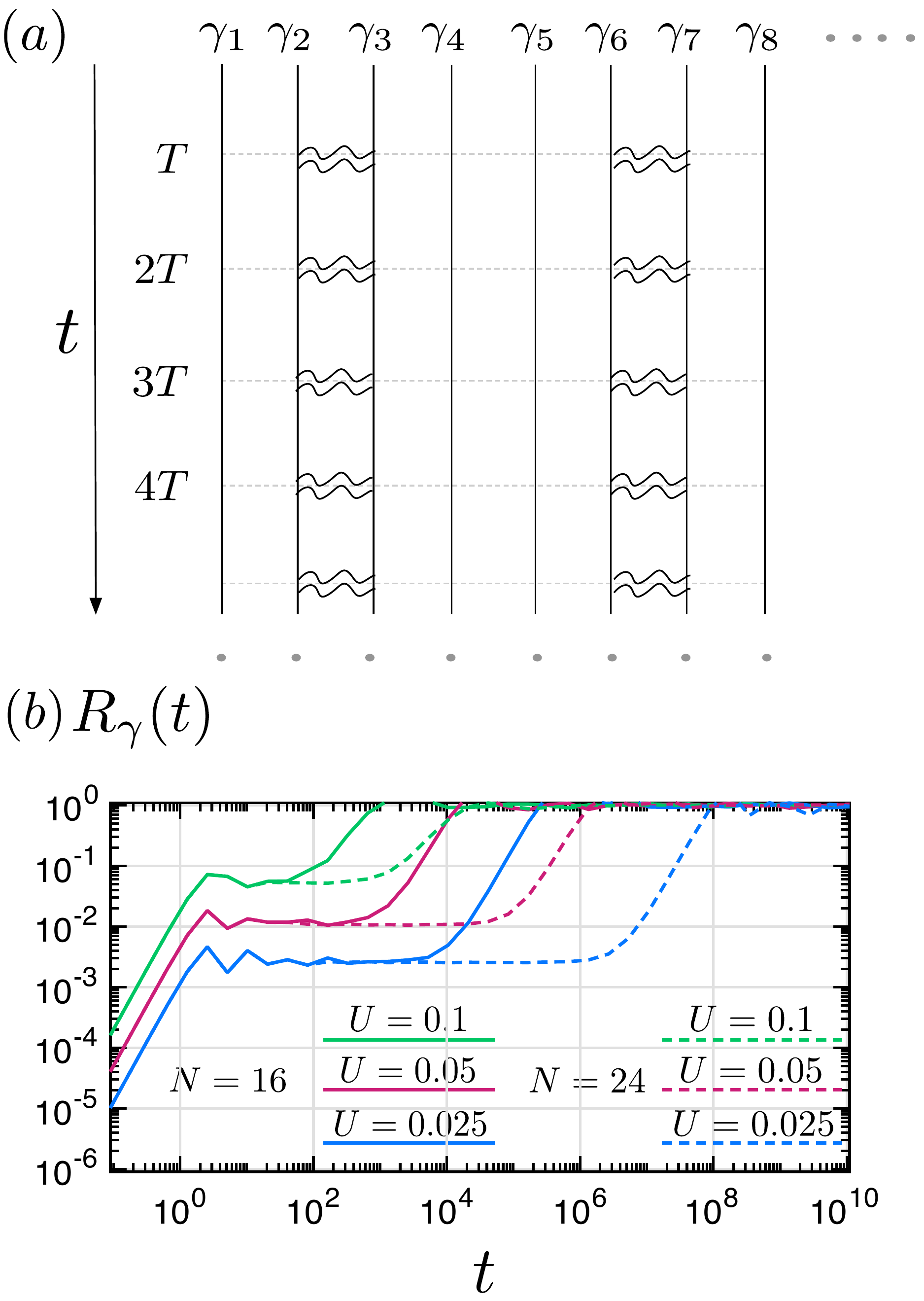}
\caption{(a) protocol to create Kitaev chain for $n_f = 1$. (b) Relaxation of the mojorana zero mode occurs on an effective timescale determined by the overlap of the majorana modes at the end of the chain. This overlap is approximately given by $1/U^{N/4}$, for $N$ majoranas.}
\label{fig:Kit}
\end{figure}

Next, we discuss particular examples for generating one-dimensional topological phases given a set of hybridized interacting majoranas. Broadly speaking, non-interacting one-dimensional systems of spinless fermions can exhibit various topological phases that are characterized by the number $n$ of decoupled majoranas at the ends of the chain~\cite{kitaev2009periodic,fidkowski2010effects}. One may view these phases as being topologically equivalent to $n$ decoupled Kitaev wires, each with a majorana at each end, protected by $n$ independent $\mathcal{Z}_2$ symmetry operators associated with the fermion parity in each wire. When these decoupled majoranas are further allowed to interact with one another, the number of independent topological phases is limited to $8$\ref{fidkowski2010effects}. In what follows, we will give examples of how one can generate two particular phases in this litany of topological phases---the simple Kitaev chain, which corresponds to the presence of one decoupled majorana at each end of the system, and the $\mathcal{Z}_2 \times \mathcal{Z}_2$ symmetry-protected phase, which has a degenerate spin-1/2 system at each end (effectively equivalent to two decoupled majoranas at each end). 

\subsubsection{1d Kitaev phase}

Let us first examine the protocol for creating a Kitaev wire from a chain of  overlapping interacting majoranas. In the starting Hamiltonian, all nearest neighbor majorana overlaps are of the same order as one another, assuming the majoranas are roughly equal spaced, and decay exponentially with inter majorana distance.  To arrive at the Kitaev phase requires that the overlaps between majoranas $\left(\gamma_{2 i + 1},  \gamma_{2 i + 2} \right) $ are weaker than the overlaps between majoranas $ \left(\gamma_{2 i}, \gamma_{2 i +1} \right)$, for integer $i$. Then, assuming that the global fermion parity $P = \prod_i \gamma_i$ is conserved (which we assume here a priori), and interactions are not too strong compared to the  majorana overlaps, the system is in the Kitaev phase, and the localization length of the decoupled majorana edge modes is set by the ratio of the strong/weak bonds. This alternating bond suppression can be achieved by the braiding operation described in Fig.~\ref{fig:Kit} (a) where majoranas $\left(\gamma_{4 i + 2}, \gamma_{4 i + 3} \right) $ are braided at period $T_0$ (for $n_f = 1$; for $n_f = 2$, for instance, the same operation would be performed at periods $2T_0$ as well, and so on).

The numerical demonstration of the feasibility of realizing the Kitaev phase using the protocol is shown in Fig.~\ref{fig:Kit} (b). We study a majorana chain with short range tunneling (overlaps decreasing by $1/e$ every lattice site), and smaller nearest-neighbor quartic interaction terms of strength $U \lesssim 1$. The Hamiltonian does not contain parity-violating terms as is natural for a physical majorana system. In Fig.~\ref{fig:Kit}, we examine the relaxation rate of the majorana at the left end of this chain, upon driving with the above protocol with $T_0 = 0.01$. The relaxation rate is probed by measuring $R_\gamma (t) = 1 - \avg{\gamma_1 (t) \gamma_1 (0)}$ (where the average is over all possible states in the Hilbert space) which begins at $0$, and rises to $1$ as the majorana decoheres with time. 

The system can be seen to be in the Kitaev phase by noting the exponential dependence of the majorana relaxation rate on the system size. More concretely, with the braiding procedure, the effective Hamiltonian can be thought of comprising two approximately decoupled Hamiltonians---one for unbraided majoranas, and another one for braiding ones. The unbraided set of majoranas tend to form a topological Kitaev chain (the effective coupling between the majoranas $\gamma_1, \gamma_4$ is weaker than that between $\gamma_4, \gamma_5$ due to physical proximity, and so on), with effective weak bonds across pairs of braided majoranas. The effective tunnel coupling between majoranas across these weak bonds is determined by $U$ (the braided majoranas require an energy of $\mathcal{O} (1)$ to be broken, and can be assumed to be effectively frozen). Consequently, the relaxation (decay) of the edge majoranas in this phase occurs due to the coupling between them, and must scale exponentially with the system size. In Fig.~\ref{fig:Kit}, we find indeed that the relaxation rate is well approximated by $U^{cN}$ (with $c \approx 1/4$), consistent with $U$ playing the role of the weak-bond hybridization in the Kitaev model.

[The initial rapid change of $R_\gamma (t)$ in Fig.~\ref{fig:Kit} is due to the ``basis rotation" between the majorana eigenmodes of the effective Hamiltonian (spread over several sites near the edge)  and the majorana basis states (strictly localized on single  edge site)].

\subsubsection{$\mathcal{Z}_2 \times \mathcal{Z}_2$ SPT phase}
\begin{figure}[htp]
\includegraphics[width=3.1in]{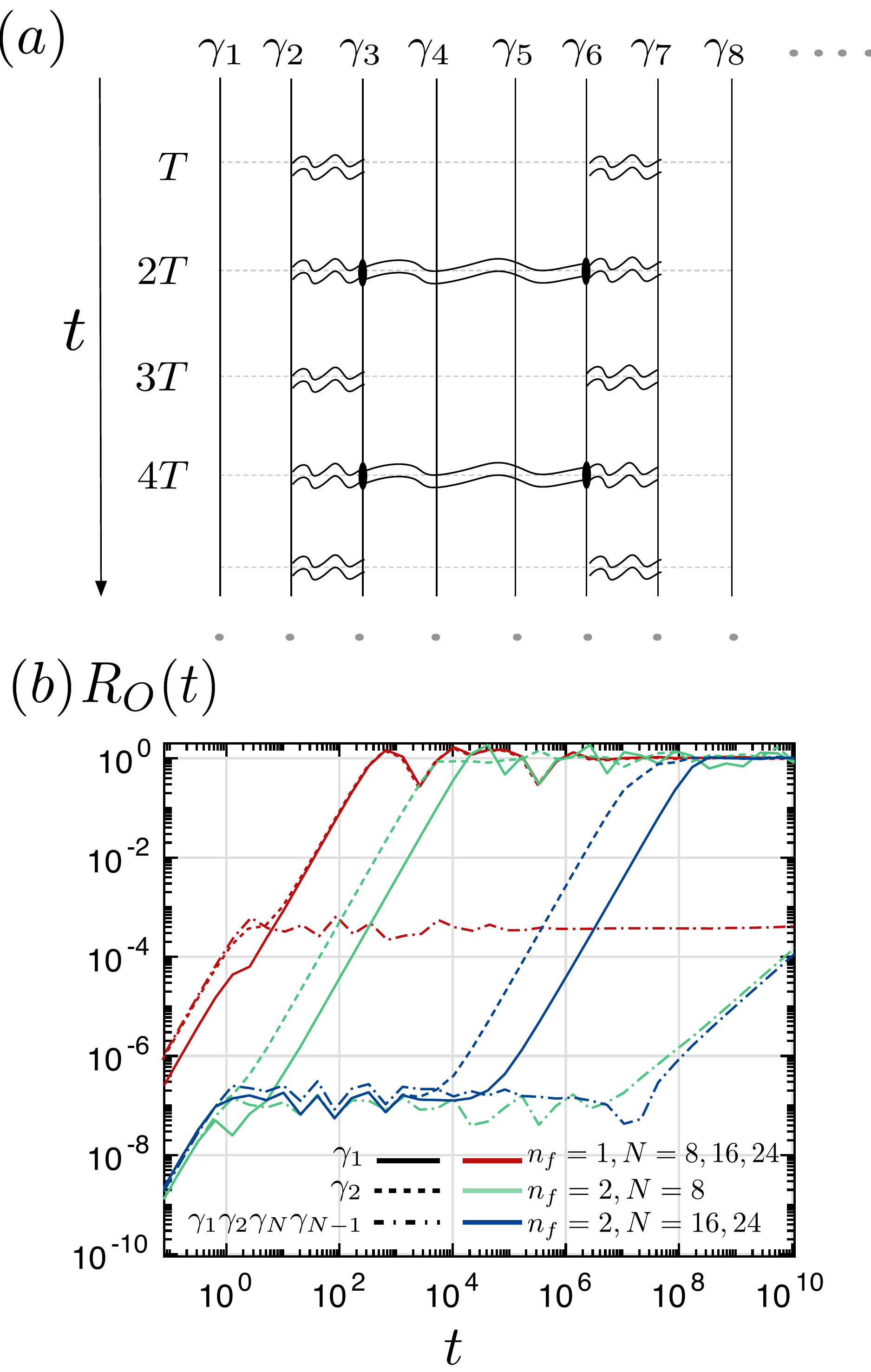}
\caption{ a) $n_f = 1$ scheme for creating $\mathcal{Z}_2 \times \mathcal{Z}_2$ SPT phase. The three draids shown at $t = 2T_0, 4T_0, ...$ may be combined into a single draid of $\gamma_2, \gamma_7$. (b) Relaxation of operators $O = \gamma_1, \gamma_2, \gamma_1 \gamma_2 \gamma_{N} \gamma_{N-1} $ is plotted. Relaxation of individual edge majoranas can occur due to i) local symmetry breaking terms which hybridize majoranas on the same edge, ii) tunneling through the bulk and hybridizing with majoranas across the other end, or iii) hybridizing with the bulk. The lack of relaxation of the parity of the zero modes, $\gamma_1 \gamma_2 \gamma_N \gamma_{N-1}$, in all cases, shows that relaxation into the bulk is absent. For $n_f = 1$, and system sizes $N = 8, 16, 24$, the edge majoranas appear to relax by hybridizing locally since the relaxation is essentially independent of system size (the plots for different $N$ largely overlap and we plot only a representative plot for clarity of presentation). For $n_f = 2$, the local symmetry breaking terms are further suppressed. Thus we see relaxation being dominated by hybridization across the wire, as seen in the dramatic system size dependence of the relaxation of edge majoranas between $N = 8, 16$. For $N \ge 16$, clearly local hybridization is again most relevant and the relaxation rate is independent of $N$.}
\label{fig:Z2SPT}
\end{figure}

We now describe the protocol for obtaining the $\mathcal{Z}_2 \times \mathcal{Z}_2$ symmetry-protected topological phase. We engineer this phase by using the logic presented in the beginning of this section. In particular, the protocol used to create the Kitaev chain can be thought of as effectively breaking the majorana chain into two independent chains, where one of the chains is in a topologically trivial phase, and the other chain is in the Kitaev phase. Concretely, the chain in the Kitaev phase consists of majoranas $\left( \gamma_1, \gamma_4, \gamma_5, \gamma_8, \gamma_9, \gamma_{12}, \gamma_{13}, ...\right) $, while the chain in the trivial phase consists of majoranas $\left( \gamma_2, \gamma_3, \gamma_6, \gamma_7, ... \right)$. In both chains, the effective bond strengths alternate between even and odd bonds due to the spatial separation of majoranas, but in the first chain, it is the even bonds that are stronger, thus rendering the chain in the Kitaev phase, while in the second chain, it is the odd bonds that are stronger.

One way of realizing the $\mathcal{Z}_2 \times \mathcal{Z}_2$ symmetry-protected phase, and the approach we take, is to convert the chain in the trivial phase into the Kitaev phase using the same protocol as described above. In particular, for the second set of braiding operations, $X_2$ applied at periods $2 T_0$, we braid pairs of majoranas in the trivial sub-chain to render this chain in the Kitaev phase, as seen in Fig.~\ref{fig:Z2SPT} (a). In the topological phase, we anticipate that the majoranas $\gamma_1, \gamma_2$ are effectively decoupled from the rest of the system. However, the symmetry-protected topological phase is more fragile in that its stability relies on the presence of symmetries preventing local hybridization between the edge majoranas $\gamma_1, \gamma_2$. In Fig.~\ref{fig:Z2SPT}, we examine the efficacy of engineering a $\mathcal{Z}_2 \times \mathcal{Z}_2$ SPT phase. In particular we examine the relaxation of the following operators: $\gamma_1$, $\gamma_2$, which are two majorana zero modes at one edge of the wire (up to order $\mathcal{O} \left( T_0^{n_f} \right)$ corrections), and $\gamma_1 \gamma_2 \gamma_N \gamma_{N-1}$ which is the parity of all the zero modes in the SPT phase (again up to $\mathcal{O} \left( T^{n_f}_0 \right)$ corrections). The relaxation of the latter operator can only happen if the zero modes are not established and can decay freely into the bulk; in all cases we see strong evidence that such a decay does not happen on the exponentially long time scales numerically probed, see Fig.~\ref{fig:Z2SPT} (b). This leaves two mechanisms for the relaxation of the majorana zero modes---hybridization with one another on the \textit{same} edge, which occurs if the symmetry protecting the SPT phase is not robust, or hybridization across the wire, which occurs due to tunneling through the gapped bulk, and is thus exponentially small in the system size. In Fig.~\ref{fig:Z2SPT}, we see evidence of both kinds of relaxation processes, with local hybridization of edge majoranas dominating at large enough system sizes. The system size at which local hybridization becomes important grows as the number of fractal layers $n_f$ grows, which further illustrates how the addition of more layers in the polyfractal protocol helps engineer better protected symmetries in the system.

\section{Experimental Implementation}
\label{sec:Hexp}

\begin{figure}[htp]
\includegraphics[width=3.3in]{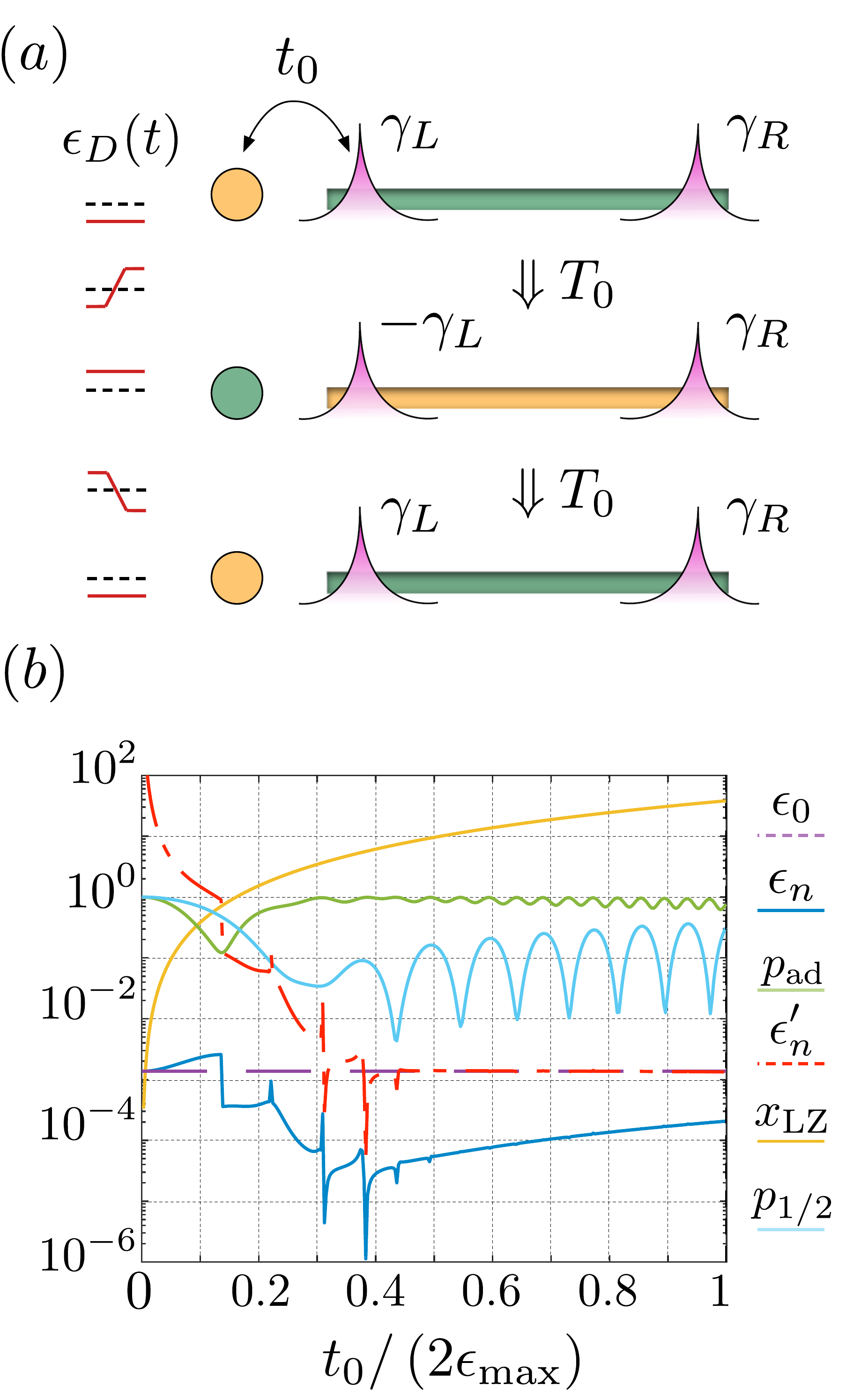}
\caption{Scheme to improve the zero bias conductance peak in quantum wires. (a) The quantum dot energy $\epsilon_D (t)$ is varied between $\left( -\epsilon_{\text{max}}, \epsilon_{\text{max}} \right)$, with $\epsilon_{\text{max}} > \epsilon_0$, the hybridization energy of the left and right majoranas $\gamma_L$ and $\gamma_R$. This is done periodically in time, with period $T_0$, each time causing the parity of the superconducting wire and the dot to flip. (b) The static hybridization $\epsilon_0$, dynamically modified hybridization energy $\epsilon_n$, and scaled hybridization $\epsilon'_n = c \epsilon_n (t/t_0)^2$ ($c = 0.37$) are varied with the dot-wire coupling $t_0$. The probability for an electron to remain in the dot over $2T_0$, $p_{\text{ad}}$, is also plotted and seen to be $1$ either in the diabatic ($x_{\text{LZ}} \ll 1$) or adiabatic ($x_{\text{LZ}} \gg 1$) limit of the Landau-Zener transitions; also plotted is the probability $p_{1/2}$ for the quantum dot electron to tunnel into the wire at time $T_0$. For $t_0/(2 \epsilon_{\text{max}}) \gtrsim 0.3$, $p_{\text{ad}} \approx 1$ and $p_{1/2} \ll 1$, corresponding to the adiabatic tunneling picture mentioned above; here we further see the scaling ansatz for $\epsilon'_n$ works remarkably well. The specific parameters associated with this simulation are provided in the main text.}
\label{fig:exp}
\end{figure}

In the above discussion, we have chosen to eschew delving into concrete realizations in view of the relative nascency of experimental work on using majoranas to engineer complex Hamiltonians. To make connection with existing experiments, we now discuss a concrete example of how the protocols of the kind proposed above may be used to improve identification of majorana modes and possibly improve their stability for the purposes of quantum operations.

The setup we consider is illustrated in Fig.~(\ref{fig:exp}) (a). It consists of a quantum dot (QD) located on one side of a superconducting wire hosting imperfect (weakly hybridized) majorana modes; the latter has been claimed to be achieved in several experiments; see Ref.~\cite{lutchyn2018majorana} for a review. Our goal is to reduce the hydridization $\epsilon_0$ that exists between the majoranas $\gamma_L$ and $\gamma_R$ at opposite ends of this wire. Of course, draiding these two majoranas induces a $\pi$-phase on both of them and thus does not lead to dynamical reduction of the hybridization between them. In order to reduce their hybridization using methods discussed in this work, we need to introduce an additional pair of majoranas, one of which can be then be braided periodically with, say, $\gamma_L$ to induce a \emph{relative} $\pi$-phase between $\gamma_L$ and $\gamma_R$, and result in a reduced effective hybridization energy $\epsilon_n < \epsilon_0$. In this setup, we model this extra pair of majoranas in terms of a single fermionic mode of a quantum dot which is coupled via a weak tunnel coupling $t_0$ to the Kitaev wire. 

Note that a draid between a majorana comprising the QD and $\gamma_L$ changes the parity of both the wire and the dot. The simplest way to achieve such parity changes is to consider modulating the local QD potential $\epsilon_D (t)$ with an amplitude greater than the hybridization $\epsilon_0$, but smaller than the bulk gap in the wire, so that an electron adiabatically tunnels into the wire from the QD in time $T_0$, and back in time $2 T_0$. Here we mention an important consideration that limits $T_0$ in this setup. Thus far, we assumed that the implementation time of the draid was negligible compared to the periodic interval $T_0$ at which majoranas are braided; in that case, suppression is more effective at smaller $T_0$. However, in the specific case we now consider, these draids are implemented using adiabatic Landau-Zener transitions between the QD and the wire, the effectiveness of which improves at slower speeds. Thus, here we cannot make $T_0$ arbitrarily short, as it limits the ramp time $T_r$ associated with the Landau-Zener ramp on the quantum dot. 

\subsection{Numerical Verification}

We now model the above using a simple numerical calculation, and show that the results are in agreement with a qualitative calculation of the dynamically reduced hybridization energy $\epsilon_n$. The numerical model we study is described by the following Hamiltonian---

\begin{align}
H &= H_w + H_D + H_{wD} \nonumber \\
H_w &= \sum_{i=1}^{L} -t c^\dagger_i c_{i+1} - \Delta c^\dagger_i c^\dagger_{i+1} - \mu c^\dagger_i c_i + \text{h.c.} \\
H_D &= \epsilon_D (t) c^\dagger_D c_D \\
H_{wD} &= -t_0 c^\dagger_D c_1 + \text{h.c.}
\label{eq:Hwire}
\end{align}
 where $\epsilon_D (t=0) = - \epsilon(t = T_0) = \epsilon_{\text{max}}$, and there's an interval of length $T_r$ in which $\epsilon_D (t)$ switches linearly between these two extremes. For subsequent times, $\epsilon_D (t)$ can be derived using $\epsilon_D (t + T_0) = - \epsilon_D (t)$. The protocol is obviously periodic with period $T_f = 2T_0$. In Fig.~\ref{fig:exp} (b), we see the effect of such driving for $t = 2, \Delta = 2, \mu = 3.6, \epsilon_{\text{max}} = 0.24$; further, $T_0 = 6 \times 10^{2}$ which is a large timescale, but satisfies $T_0 \epsilon_0 \lesssim 1$, and is thus short compared to the static hybridization, but otherwise as long as one possibly can get. Finally, the ramp time is $T_r = T_0/2$. We will justify these choices below. 

We simulate the model by Trotterizing and numerically computing the Floquet unitary corresponding to the protocol. To find the dynamically suppressed hybridization energy,  we search for the eigenmode with the largest inversee participation ratio (IPR) on the wire sites. Finally, the quasi-energy of this mode is used to determine the effective hybridization energy $\epsilon_n$. The fact that these Floquet modes correspond to the original majorana modes can be verified as well by directly visualizing the weight of these modes on wire sites within the period $2 T_0$; see Appendix~\ref{app:floquetmaj}. 

In this particular case, we find a best case suppression of about $100$ times compared to the original hybridization between the majoranas, which occurs at smallest values of the QD-wire coupling at which the Landau-Zener transitions become nearly adiabatic, ensuring desired parity flipping on every passage of the QD level $\epsilon_D(t)$ through zero. This hybridization is well explained by an analytical argument which we turn to next. 

\subsection{Analytical argument}

We now present a simplified description of the above numerical model and explain our findings, in particular, what determines the extent $\epsilon_n / \epsilon_0$ of the possible suppression of the static hybridization between the marjorans due to this dynamical scheme. 

Assuming all dynamical processes occur on time scales slow compared to the bulk gap of the wire, we can work in the low-energy subspace of the edge majorana modes and the quantum dot. This is a $4$-dimensional subspace is further reduced into two fixed total parity (of both wire and QD) subspaces of dimension $2$ each. If we work in the occupation basis of the fermionic mode constructed from $\gamma_L, \gamma_R$, and the fermionic mode of the QD, the $2$-by-$2$ Hamiltonian corresponding to these sectors is given by 

\begin{align}
H^{\pm} &= \frac{1}{2} \epsilon_D (t) +  \begin{pmatrix} \mp \epsilon_0 - \frac{1}{2} \epsilon_D (t) & \frac{t_0}{\sqrt{2 \xi}} (1 \pm i \beta) \\ \frac{t_0}{\sqrt{2 \xi}} ( 1 \mp i \bar{\beta} ) & \pm \epsilon_0 + \frac{1}{2} \epsilon_D (t)  \end{pmatrix}
\end{align}

where $H^+$ acts on the basis of even parity states $\ket{1_D 1_w} , \ket{0_D 0_w}$, and $H^-$ acts on the basis of odd parity states $\ket{0_D 1_w} , \ket{1_D, 0_w}$. Here $\xi$ is the localization length of the edge majoranas.

The diagonal elements of $H^\pm$ follow straightforwardly by tracking the local poential $\epsilon_D (t)$ on the QD and the energy of the fermionic mode $\epsilon_0$ constructed from $\gamma_L, \gamma_R$. The off-diagonal matrix element is naturally proprtional to $t_0$; below we justify its precise form by arriving at the low-energy description of the operator $c_1, c^\dagger_1$ that occurs in the tunneling matrix element---

\be
c^\dagger_1 \rightarrow \frac{1}{\sqrt{\xi}} (\gamma_L + \beta \gamma_R ) = \frac{1}{\sqrt{2}} \left( c_0 ( 1 + i \beta) - c_0^\dagger ( 1 - i \beta) \right)
\ee

and similarly for $c_1$. Here we assumed that, in the low-energy limit, only the decomposition of $c_1, c^\dagger_1$ in terms of the low-energy majorana modes $\gamma_L, \gamma_R$ needs to be considered, $\gamma_L$ has weight $\sim 1/\sqrt{\xi}$ at the first site, and the result is finally expressed in terms of the fermionic operators $ c_0 = (\gamma_L + i \gamma_R)/\sqrt{2}$. Here $\beta$ is some complex number of small amplitude $\sim e^{-L/\xi}$, where $L$ is the wire length; this reflects the fact that the right majorana is localized at the opposite end of the wire to the tunneling operator $c_1$. 

Now, the protocol is expected to be effective in the adiabatic regime wherein the wire's parity switches periodically. In this regime, the Floquet eigenstates are expected to be closely related to the instantaneous eigenstates of $H^{\pm}$ and the effective hybridization between the majoranas is given by the time-averaged dynamical phase difference between the ground (or excited) states of $H^+$ and $H^-$, which correspond to different parity sectors in the wire far away from the Landau-Zener transition regions. This follows analogously to how the hybridization energy in the static setting can be ascertained from the dynamical phase accrued in time between ground states of different parities. In particular, we find

\begin{align}
\epsilon_n &= \frac{1}{2T_0} \abs{\int_0^{2 T_0} dt \;\left[E^+ (t) - E^- (t)\right]}, \; \; \text{where} \nonumber \\
E^{\pm} &= \frac{\epsilon_D (t)}{2} - \sqrt{ \left( \pm \epsilon_0 - \frac{\epsilon_D (t)}{2} \right)^2 + \frac{t^2_0}{2 \xi} \abs{1 \pm i \beta}^2 }    
\label{eq:Epm}
\end{align}
are the energies of the ground states of $H^\pm$. If $\beta$ was negligible, or purely real, using $\epsilon_D (t) = - \epsilon (t + T_0)$, one can show that $\epsilon_n = 0$. One can estimate $\epsilon_n$ by noting that the phase difference is primarily accrued when the tunneling matrix element is comparable to the energy. This occurs for two periods of time $\sim \frac{t_0 T_r}{ \epsilon_{\text{max}} \sqrt{\xi}} $, and the rate of accruing phase during this time is $\sim \frac{t_0 \abs{\beta}}{\sqrt{\xi}}$. Further estimating $\epsilon_0 \sim t e^{-L/\xi} \sim t \abs{\beta}$, we find

\be
\epsilon_n \sim \frac{t_0 \abs{\beta}}{\xi} \cdot \frac{t_0 T_r}{\epsilon_{\text{max}} T_0} \Rightarrow \frac{\epsilon_n}{\epsilon_0} \sim \frac{1}{\xi} \cdot \frac{T_r}{T_0} \cdot \frac{t^2_0}{t \epsilon_{\text{max}}} 
\ee

We again emphasize this result is valid only when the parameter $x_{\text{LZ}} = \frac{t^2_0 T_r}{\hbar \xi \epsilon_{\text{max}}}$ that controls the adiabaticity of the Landau-Zener transitions is significantly greater than $1$. In this regime, the numerically determined variation of the effective $\epsilon_n$ with the QD-wire coupling $t_0$ agrees remarkably well with these arguments; see $\epsilon'_n$ in Fig.~\ref{fig:exp} (b). Finally, using the criterion $x_{\text{LZ}} \gtrsim 1$ implies that the most suppression achievable in this protocol is given by 

\be
\frac{(\epsilon_n)_{\text{min}}}{\epsilon_0} \sim \frac{\hbar/T_0}{t} \gtrsim \frac{\epsilon_0}{t}. 
\label{eq:majsupp}
\ee

In summary, the best suppression is achieved for $x_{\text{LZ}} \gtrsim 1$, and \emph{largest} feasible drive period $T_0 \sim 1/\epsilon_0$ determined by the original majorana hybridization. Thus, the protocol works better when one already has reasonably well localized majorana modes, but depending on the various experimental setups, this may represent a few orders of magnitude of suppression from the static hybridization.  

\subsection{Experimental feasibility, and other remarks.}

Experimentally, the hybridization between majoranas manifests in the shift of the density of states (DOS) resonance away from zero energy. Larger spread of the majorana wavefunction into the bulk correspond to a larger overlap between edge majoranas, and thus to the larger deviation of the DOS peak from zero energy.
The peak in DOS is detectable via local tunneling, either from  normal~\cite{mourik2012signatures,albrecht2016exponential,nadj2014observation} or superconducting lead/tip~\cite{ruby2015tunneling}.

The hybridization between majoranas can be suppressed using the protocol described above. Consequently, if we now measure low frequency tunneling into the wire, we should observe a sharp peak much closer to zero bias, even if the static hybridization between the majoranas was significant. The measurement has to be sufficiently slow -- i.e., the electron tunneling rate has to be below the frequency of the protocol (with the manipulation protocol being faster than the static hybridization between majoranas), but slower than the inverse superconducting gap. 

In present experimental setups consisting of semiconductor wires or magnetic adatom chains, the induced superconducting gap is typically of the order $\text{meV}$. The majoranas are delocalized over a significant fraction of the entire wire (of length $\sim 100 nm - 1\mu \text{m}$), and consequently the zero bias peak is observed at voltages corresponding to $10-100 \mu \text{eV}$.  Thus, we anticipate that by driving at a frequency $1/T_0 \sim 10-100 \; \mu \text{eV}/ \hbar \approx 50 \text{MHz} - 0.5 \text{GHz}$ can achieve a suppression of hybridization by about two orders of magnitude [using Eq.~(\ref{eq:majsupp})]. Following our numerical experiment, the quantum dot, engineered using an appropriate set of potential gates, would have to be manipulated by an oscillating voltage of amplitude $\sim 0.1$ meV, along with engineering a comparable tunnel coupling. These numbers appear to be reasonable and within experimental reach.    

We note that in this work, we have described a particular protocol to suppress majorana hybridization. There may be more elaborate protocols consisting of multiple Landau-Zeneer transitions which further improve on the best case suppression found in Eq.~(\ref{eq:majsupp}). We leave discussion of more elaborate protocols to future work. 

Here we examined dynamical suppression of hybridization in a regime where a large majorana localization length $\xi$ arises due to the chemical potential $\mu$ being close to the band edge, $\mu \lesssim 2 t$. Instead, if we examine the case of perfect particle hole symmetry---$\mu = 0$---the suppression is infinitely strong, that is, $\epsilon_n = 0$ exactly. This is a very fine tuned result whose discussion we leave to Appendix~\ref{app:mu0}. We additionally provide some results regarding the Floquet dynamics of the dynamically engineering majoranas. 

\section{Discussion and Conclusions} 

In this work, we  theoretically explored a possibility of using braiding operations on majoranas to engineer novel Hamiltonians. Our proposal is based on the topological robustness of the $\pi-$phase acquired by Majoranas in otherwise trivial draiding (double-braiding) operations. Repeated application of such braids can be stable to heating if the braiding is performed at sufficiently high frequencies, by way of pre-thermalization, or by introducing disorder, by way of many-body localization. As we show, the phases accrued by such repeated braiding can then be used to selectively eliminate certain terms in the Hamiltonian, relatively enhancing others, they can be used to engineer additional symmetries and thus realize  symmetry protected topological phases, and further, to stabilize quantum bits. 

Finally, we discussed how such a protocol can be used to suppress the deleterious hybridization between edge majoranas in experimentally relevant finite Kitaev wires. In this protocol we substitute the hard to engineer braiding operations with adiabatic Landau Zener tunneling between pairs of majorana modes. The adiabaticity requirement implies the protocol cannot be performed arbitrarily fast but we estimate that our protocol can be used to suppress majorana hybridiation in current experimental setups by a few orders of magnitude. 

\section{Acknowledgements}

We would like to acknowledge discussions with Dima Pikulin, Bela Bauer, Torsten Karzig, Roman Lutchyn, and Arijeet Pal. We are particularly grateful to Yuval Oreg for inspiring us to search to a simple experimental demonstration, which lead to the proposal presented in Section  \ref{sec:Hexp}. We also acknowledge hospitality of KITP,  supported in part by the National Science Foundation under Grant No. NSF PHY-1748958, where this work was initiated.
 KA acknowledges funding from NSERC Grants RGPIN-2019-06465 and DGECR-2019-00011.
 IM was supported by the Materials Sciences and Engineering Division, Basic Energy Sciences, Office of Science, U.S. Dept. of Energy. 

\begin{figure}[htp]
\includegraphics[width=3in]{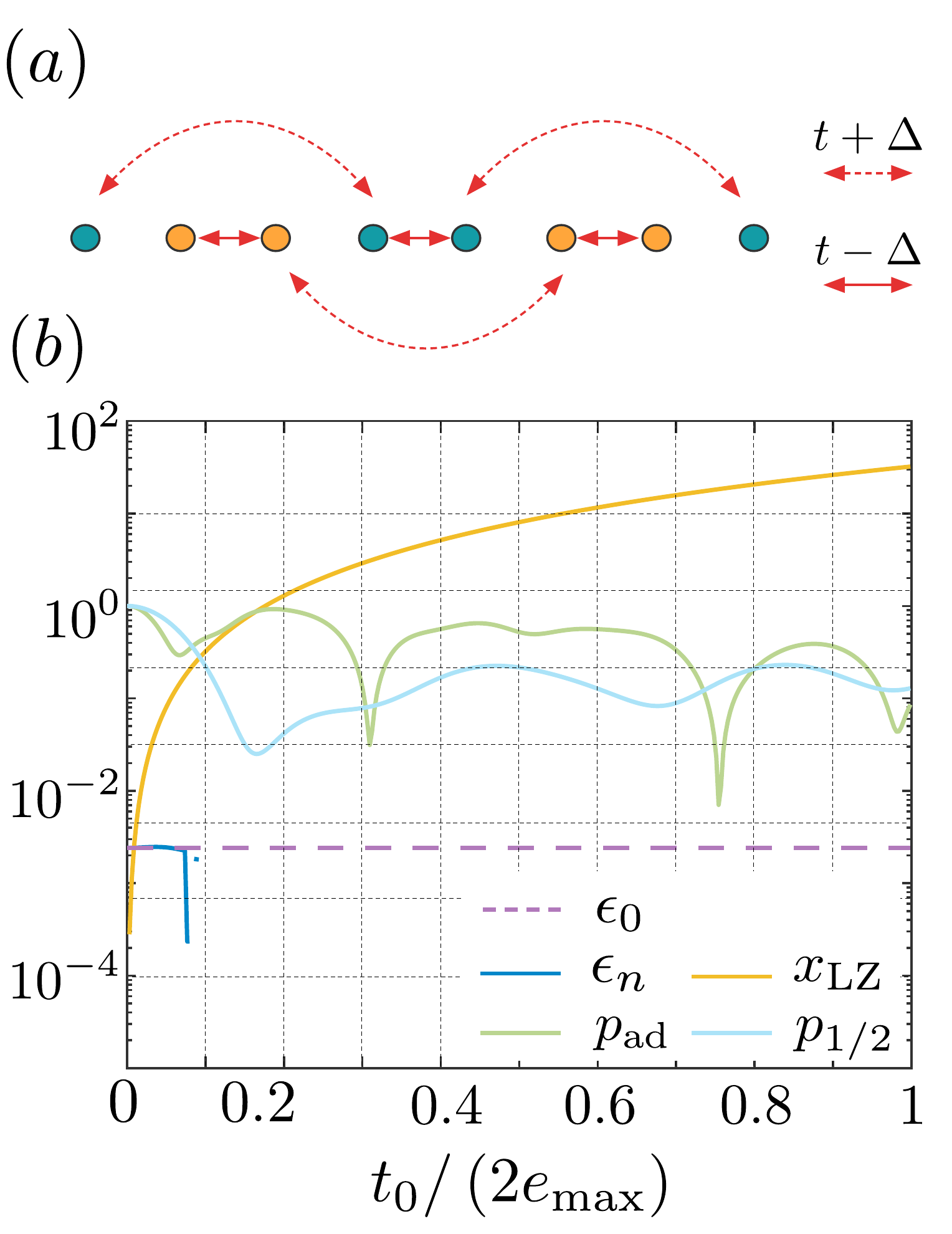}
\caption{(a) The Kitaev chain is decoupled into two chains for $\mu = 0$. For even $L$ (shown), one of these chains is in the topological phase, and one in the trivial phase. For odd $L$ both chains have an odd number of majorana modes, and thus necessarily house perfectly decoupled majoranas, although these majoranas are themselves not localized. (b) When the Landau-Zener transitions first become adiabatic, $x_{\text{LZ}} \gtrsim 1$, the dynamically suppressed hybridization drops to identically zero.}
\label{fig:app1}
\end{figure}

\appendix

\section{Perfect hybridization suppression for $\mu = 0$}
\label{app:mu0}

\begin{figure*}[htp]
\includegraphics[width=\textwidth]{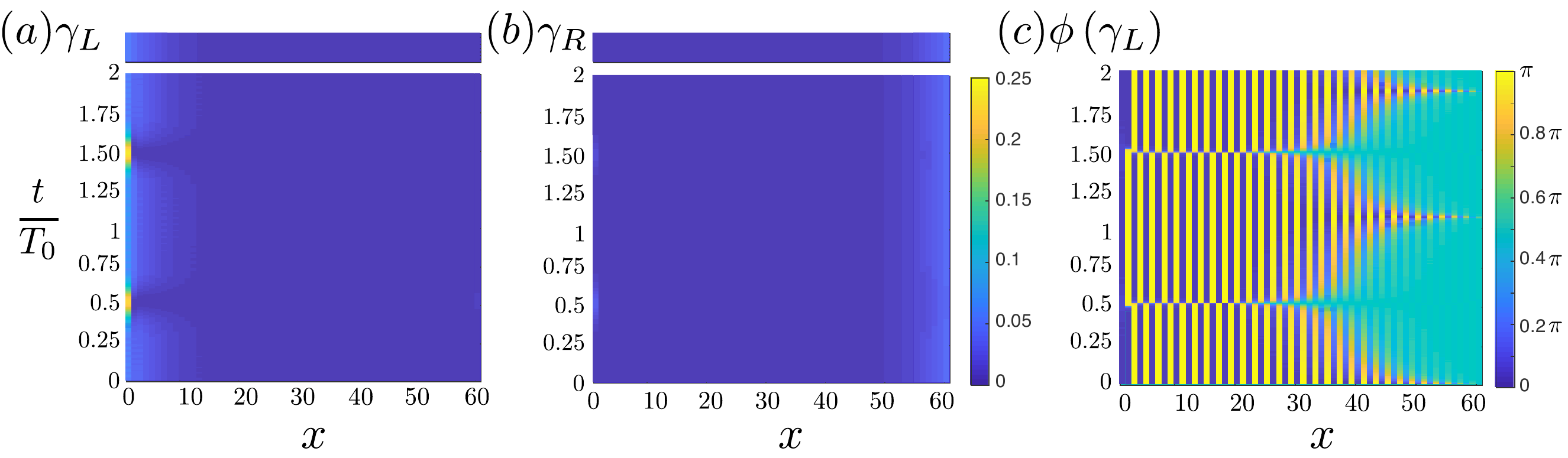}
\caption{(a) and (b): The weight of the left and right Floquet majoranas $\gamma_L, \gamma_R$ respectively on the quantum dot site (site $0$ in the plots) and the wire sites (for $L = 60$) over a single period of time $2 T_0$ of the protocol. The majoranas do not appear to change their form significantly in comparison to their static counterparts (top color bar) over most of the time evolution, except during the short intervals at which Landau Zener tunneling occurs. (c) The left majorana acquires a $\pi$  phase as anticipated after each Landau-Zener transition. The right majorana, not shown here, does not acquire any phase.}
\label{fig:app2}
\end{figure*}
For the special particle-hole symmetric case of $\mu = 0$, the majorana chain immediately decouples into two separate subchains. Let's first look at even (physical) length chains [in Fig.~\ref{fig:app1} (a), we show $L = 4$ corresponding to $8$ majoranas].  Of two subchains, only one (depending on whether $t + \Delta > t - \Delta$ or vice-versa) happens to be in the topological phase, and thus happens to house both majoranas. If tunneling occurs only into the first fermionic site in the wire, noted in Eq.~(\ref{eq:Hwire}) by $c_1, c^\dagger_1$, then the relevant tunnel coupling is only to either one of $\gamma_1$ or $\gamma_2$ (depending on which subchain is topological), and consequently, in the low-energy limit, the effective tunnel coupling occurs via either one of the operators $c_1 + c^\dagger_1$ or $-i (c_1 - c^\dagger_1)$, in which case $\beta$ is purely real and the phase difference between the even and odd parity sectors vanishes identically; see Eq.~(\ref{eq:Epm}). Thus, in this instance, we expect the hybridization to be perfectly suppressed in the dynamical protocol; this is verified numerically in Fig.~\ref{fig:app1} (b).

(We add a note of caution here that even for large Landau-Zener parameter, $p_{\text{ad}}$ appears to deviate significantly from $1$. These may merit more detailed investigation.)

Finally, we note that the case for odd length chains is even more special in that odd length chains decouple perfectly into two decoupled chains with an odd number of majoranas each. These majoranas are thus mandated to be perfectly decoupled from one another even in the static case.

\section{Dynamics of Floquet majoranas}
\label{app:floquetmaj}

We verify our adiabatic tunneling picture of the Landau-Zener transitions, and the phase flipping of the left majorana after every Landau-Zener transition (akin to draid with a quantum dot mode); see Fig.~(\ref{fig:app2}). We plot the weight of the Floquet majoranas as a function of time (over the course of one complete period of  $2 T_0$) on the quantum dot and wire sites. These majoranas are defined as the real and imaginary parts of the eigenmode of the diagonalized Floquet Unitary with the largest inverse participation ratio (IPR) on the wire sites. Due to their reality, they have equal particle/hole content on each lattice site; this weight in plotted in Fig.~(\ref{fig:app2}). We see that the majoranas clearly resemble their static counterparts except in two ways anticipated by their theoretical arguments---1) the majorana weights change during the short intervals representing Landau-Zener tunneling, 2) the left majorana acquires a $\pi$ phase after each such tunneling event, as anticipated, and desired.  

\bibliographystyle{apsrev4-1}
\bibliography{BraidingNew2}

\end{document}